\documentstyle[12pt]{article}

\newcommand{\be}{\begin{equation}}
\newcommand{\ee}{\end{equation}}
\newcommand{\ba}{\begin{eqnarray}}
\newcommand{\ea}{\end{eqnarray}}
\newcommand{\la}{\lambda}

\newcommand{\Tr}{\rm Tr}
\newcommand{\tr}{\rm tr}
\newcommand{\g}{\gamma}
\newcommand{\e}{\epsilon}

\begin{document}
\hoffset=-.4truein\voffset=-0.5truein
\setlength{\textheight}{8.5 in}
\begin{titlepage}
\hfill{LPTENS 07-037}
\begin{center}

\vskip 0.6 in
{\large   Intersection theory from duality and replica}
\vskip .6 in
\begin{center}
{\bf E. Br\'ezin$^{a)}$}{\it and} {\bf S. Hikami$^{b)}$}
\end{center}
\vskip 5mm
\begin{center}
{$^{a)}$ Laboratoire de Physique
Th\'eorique, Ecole Normale Sup\'erieure}\\ {24 rue Lhomond 75231, Paris
Cedex
05, France. e-mail: brezin@lpt.ens.fr{\footnote{\it
Unit\'e Mixte de Recherche 8549 du Centre National de la
Recherche Scientifique et de l'\'Ecole Normale Sup\'erieure.
} }}\\
{$^{b)}$ Department of Basic Sciences,
} {University of Tokyo,
Meguro-ku, Komaba, Tokyo 153, Japan. e-mail:hikami@dice.c.u-tokyo.ac.jp}\\
\end{center}
\vskip 3mm 
{\bf Abstract}
\end{center}
Kontsevich's work on Airy matrix integrals has led to explicit results for the
intersection numbers of the moduli space of curves. In this article we show  that a duality between k-point functions on $N\times N$ matrices and N-point functions of $k\times k$ matrices, plus the replica method, familiar in the theory of disordered systems, allows one to recover  Kontsevich's results on the intersection numbers, and to generalize them to other models. This provides an alternative and simple  way to compute intersection numbers with one marked point, and leads also  to some new results. 
\end{titlepage}
\vskip 3mm
 
\section{Introduction}

After Witten's celebrated conjectures \cite{Witten, Witten2} on the relation between intersection numbers on moduli spaces of curves and the KdV hierarchy, and Kontsevich's proof \cite{Kontsevich},  the literature on the subject, both from the point of view of mathematics or from its string theory relationship,  has become considerable.  We want to add here a new method, duality plus replica,  which allows one to recover easily some results which are difficult to obtain by fancier methods, and provides some new results as well. 

In a previous article \cite{BH0} we have used 
explicit   integral representations for the correlation functions  \cite{BHa,BHb,BHc} for a Gaussian unitary ensemble (GUE) of random matrices $M$ in the presence of an external matrix source.  The probabililty distribution for $N\times N$ Hermitian matrices is 
\be\label{PA}
P_A(M) = \frac{1}{Z_A} e^{-\frac{N}{2}{\tr} M^2 - N {\tr} M A}
\ee
From this representation we have obtained  the correlation functions of the 'vertices' 
\be
V(k_1,...,k_n) = \frac{1}{N^n}<{\tr} M^{k_1} {\tr} M^{k_2} \cdots {\tr} M^{k_n}>
\ee

As is well-known from Wick's theorem (when the source $A$ vanishes) such correlation functions are just the numbers of pairwise gluing  of the legs of the vertex operators.
The dual cells of these vertices are polygons,
whose edges are pairwise glued. This generates orientable surfaces, 
discretized
Riemann surfaces whose  genus is related to the power of $1/N^2$.
 
Okounkov and Pandharipande \cite{OP,O1,O2} 
have shown that the intersection numbers,  computed by Kontsevich \cite{Kontsevich},
may be obtained by taking
a simultaneous large $N$ and large $k_i$ limit. In our previous work we have used the exact integral representation valid for finite $N$ of those vertex correlation functions, and obtained explicitely the scaling region for large $k_i$ and large N by a simple saddle-point.  This led to a practical way
to compute  intersection numbers from a pure Gaussian model, much simpler than
Kontsevich's Airy matrix model. 

In this article we want to show that duality and replica may be used also to recover easily earlier results, to establish some new ones and give support to Witten's conjecture. In order to make this article self-contained we will add appendices in which we rederive some of the steps leading to the representation that we are using : \\(i) explicit formulae, valid for arbitrary $N$ and arbitrary source matrix $A$ for the average
\be
U(s_1,\cdots,s_k) = \langle \rm{Tr} e^{s_1M} \cdots \rm{Tr} e^{s_k M} \rangle
\ee
which rely on the Itzykson-Zuber formula \cite{Harish-Chandra, Itzykson-Zuber} (appendix A). \\
(ii) a duality representation for the average of characteristic determinants 
$$ \langle \det (\lambda_1-M) \cdots  \det (\lambda_k-M) \rangle $$
in terms of another GUE integral, but in which the random matrices are $k\times k$ , 
whereas the initial problem involved $N\times N$ matrices (appendix B) \cite{BHe}.  
(This duality seems to be a simple reflection of the open string/closed string duality \cite{Hashi}).  With the help of these two kinds of results we may proceed to the replica approach based on the simple relation 
\be\label{replica}
 \lim_{n \to 0}\   \frac{1}{n}\  \frac{\partial}{\partial \la} [\det(\la -B)]^{n}   =  \rm{tr} \frac {1}{\la -B} .
 \ee
 Therefore after applying these two steps we end up looking for an $n$ goes to zero limit on matrix integrals whose size vanishes with $n$, a very different problem from the familiar large $N$ limit. Its Feynman graph representation connects directly with Riemann surfaces with marked points of maximum genera. 

\section{The duality plus replica strategy}
\vskip 2mm
We first consider  the average of  products of  characteristic
polynomials,  
 defined as 
\ba
F_k(\lambda_1,...,\lambda_k) &=& \frac{1}{Z_N}< \prod_{\alpha=1}^k {\rm det}(\lambda_\alpha - M) >_{A,M}\nonumber\\
&=& \frac{1}{Z_N}\int dM \prod_{i=1}^k {\rm det}(\lambda_i\cdot {\rm I} - {\rm M}) e^{-\frac{N}{2}
\tr M^2 + N  \tr M A} 
\ea
where $M$ is an $N\times N$ Hermitian random matrix,  $A$ a given Hermitian matrix, whose eigenvalues are $(a_1, \cdots, a_n)$ and $ Z_N$ the normalization constant of the probability measure (for $A=0$).
We have shown earlier  \cite{BHe} that this correlation function has also a dual expression.  This duality interchanges $N$, the
size of the random matrix,  with
$k$, the number of points in $F_k$,   as well as the matrix source $A$ with the diagonal
matrix  $\Lambda = {\rm diag}(\lambda_1,...,\lambda_k)$. This duality reads \cite{BHe}
\be\label{dualB}
F_k(\lambda_1,...,\lambda_k)= \frac{1}{Z_k}\int dB \prod_{j=1}^N [{\rm det}(a_j - i B)] e^{-\frac{N}{2}\tr ( B-i\Lambda)^2}
\ee
where $\Lambda = {\rm diag}(\lambda_1,...,\lambda_k)$ and $B$ is a $k \times k$ 
Hermitian matrix. (The normalization is now on GUE ensembles of Hermitian $k\times k$ matrices $Z_k= \int dB \rm{exp}(-\frac{1}{2} \rm{tr} B^2))$. The derivation is reproduced in appendix B. 

If we specialize this formula to a source $A$ equal to the unit matrix,  a trivial shift for the original $N\times N$  matrices M, which has the effect of making the support of Wigner's semi-circle law, for the density of eigenvalues of M,  to lie between 0 and 2 rather than (-1,+1), the formula (\ref{dualB}) involves 
\ba\label{expandB}
{\rm det}(1 - i B)^N &=& {\rm exp}[ N {\tr} {\rm ln}(1 - i B)] 
\nonumber\\
&=&{\rm exp}[ - i N {\tr} B + \frac{N}{2} {\tr} B^2 + i \frac{N}{3}{\tr} B^3 + \cdots]
\ea

The linear term in $B$ in (\ref{expandB}), combined with the linear term of the exponent
of (\ref{dualB}), shifts $\Lambda$ by one.
The $B^2$ terms in (\ref{dualB}) cancel. In a scale in which the initials $\lambda_k$
are close to one, or more precisely $N^{2/3}(\lambda_k-1)$ is finite, the large $N$
asymptotics of (\ref{dualB}) is given by matrices $B$  of order
$N^{-1/3}$. Then the higher terms in (\ref{expandB}) are negligible and we are left with
terms linear and cubic in the exponent, namely
\be\label{dualB2}
F_k(\lambda_1,...,\lambda_k)= e^{\frac{N}{2} \rm{tr}\Lambda^2}\int dB   e^{i\frac{N}{3}{\tr} B^3 +
i N {\tr} B (\Lambda-1)}.
\ee 
So finally for this edge behavior problem, in which the matrix $B$ is of order $N^{-1/3}$, and $\lambda-1$ of order $N^{-2/3}$ we can rescale $B$  and $\lambda-1$ to get rid of $N$.  The result is nearly identical to 
the matrix Airy integral, namely Kontsevich's model \cite{Kontsevich}, which gives  the
intersection numbers of moduli of curves. The original Kontsevich partition function was defined as 
\be\label{Konts}
Z = \frac{1}{Z^\prime}\int dB e^{-\tr \Lambda B^2 + \frac{i}{3}\tr B^3}
\ee
where $Z^\prime = \int dB e^{-\tr B^2}$. The shift $B \to B +i\Lambda$,
eliminates the ${B^2}$ term and one recovers (\ref{dualB2}) up to a trivial rescaling. 

Let us apply this to a one-point function.  In view of the replica limit we specialize those formulae to
\be\label{r0} < [{\rm det}(\lambda - M)]^n >_{A,M} =<   [{\rm
det}(1 - i B)]^N >_{\Lambda,B}
\ee
where $B$ is an $n\times n$ random Hermitian matrix. For this edge problem we have chosen for the source matrix $A$, the N-dimensional unit matrix, whereas    $\Lambda $  is a
multiple of the
$n\times n$ identity  matrix :
$ \Lambda = {\rm diag}(\lambda,...,\lambda)$. Note that the average in the l.h.s. of (\ref{r0}) is meant for the $N\times N$ GUE ensemble, whereas in the r.h.s. the average is performed on $n\times n$ Hermitian matrices with the weight (\ref{dualB2}) or (\ref{Konts}).

The strategy that we will used is thus to take the integral (\ref{Konts}) for $\Lambda = \lambda \times\mathbf {1}$ and  expand it in powers of the cubic term. The formulae that will be established, being exact for finite $n$,  are easily continued in $n$ and allow one to take the $n\to 0$ limit. The method relies on explicit exact representations of Gaussian averages \cite{BHc, Kazakov} in the presence of an external matrix source (in appendix A the main steps of the derivation have been recalled).  As a result we will obtain formulae for quantities such as $$ \lim_{n \to 0} \frac {1}{n} <(\rm{tr} B^k)^{l}>$$ 
computed with the (normalized) Gaussian weight $Z^{-1}\rm{exp}(-\frac{1}{2}\rm{tr} B^2)$ on $n\times n$ Hermitian matrices ;  thereby this procedure gives the values of the intersection numbers of the moduli of curves for a number of cases. 

\section{Evolution operators and replica limit}
We will rely on explicit expressions for 
\be  
U_A(s_1,\cdots,s_k) = \frac{1}{n}\langle \rm{tr} e^{s_1B} \cdots \rm{tr} e^{s_k B} \rangle 
\ee
for a probability measure on $n\times n$ Hermitian matrices  in the presence of an Hermitian  matrix source A, whose eigenvalues are $a_1, \cdots, a_n$. The average is thus defined with the normalized weight
\be  P_A(B) = \frac{1}{Z} e^{-\frac{1}{2} \rm{tr} B^2 + \rm{tr} AB} \ee
Then one has (see appendix A) for the one-point function
\be U_A(s) = \frac{1}{n}\langle \rm{tr} e^{sB} \rangle = \frac{e^{\frac{s^2}{2\la}}}{ns}\oint \frac{du}{2i\pi} e^{su/\la} \prod_1^n (\frac{u-a_{\alpha} +s}{u-a_{\alpha} })
\ee
the contour, in the complex u-plane, circling around the $n$ poles $a_{\alpha}$. We have chosen the normalization $U_A(0) = 1$. 
This formula simplifies for a vanishing external source (although a non-vanishing source simplifies the derivation) to 
\be \label{1-point}
 U(s) =   \frac{e^{\frac{s^2}{2\la}}}{ns}\oint \frac{du}{2i\pi} e^{su/\la}(1 +\frac{s}{u})^n
\ee
This representation leads to a simple continuation in $n$, and to an expansion in powers of $n$. For instance it gives 
\be \label{n=0}
 \lim_{n\to 0} U(s) =   \frac{e^{\frac{s^2}{2\la}}}{s}\oint \frac{du}{2i\pi} e^{su/\la}\log{ (1+\frac{s}{u})} .\ee
This last contour integral reduces to the integral of the discontinuity of the logarithm, giving readily
\be
 \lim_{n\to 0} U(s) = \frac{\sinh {(\frac{s^2}{2\la})}}{(\frac{s^2}{2\la})}
\ee
and thus 
\be 
\lim_{n\to 0} \frac{1}{n} \langle \rm{tr} B^k\rangle = \frac{4k!}{\la^{2k} 4^k (2k+1)!}
\ee
Note that in terms of Feynman diagrams with double lines, the limit $n\to 0$ selects the diagrams with one single internal index all along the lines of the diagrams. Those diagrams correspond to a surface of maximum genus for a given number of vertices.

The same strategy works  for higher point-functions. The k-point function is given (for a vanishing source)  by
\ba 
&& U(s_1,\cdots, s_k) = \frac{1}{n} \langle \rm{tr} e^{s_1B} \cdots \rm{tr} e^{s_k B} \rangle \\ \nonumber
&& = (-1)^{k(k-1)/2} e^{\sum_1^k \frac{s_i^2}{2\la}}\oint \prod_1^k \frac{du_i}{2i\pi} e^{\sum_1^k (u_is_i/\la)} \prod_1^k (1+\frac{s_i}{u_i})^n \det \frac{1}{u_i+s_i-u_j}
\ea

Again the continuation to non-integer $n$ is straightforward and leads to 
\ba \label{k-point}
\lim_{n\to 0} U(s_1,\cdots, s_k)= 
(-1)^{k(k-1)/2} e^{\sum_1^k \frac{s_i^2}{2\la}}\\ \nonumber \times \oint \prod_1^k \frac{du_i}{2i\pi} e^{\sum_1^k (u_is_i/\la)} \sum_1^k \log{(1+\frac{s_i}{u_i})} \det \frac{1}{u_i+s_i-u_j}
\ea

The calculation of the  contour integrals is more cumbersome, but all the integration can be done explicitly to the end and give a remarkably compact result (for more details see appendix C). If we denote
\be \sigma = s_1+\cdots+s_k
\ee
they lead to
\be \label{sh}
\lim_{n\to 0} U(s_1,\cdots, s_k)  = \frac{\la}{\sigma^2}\prod_1^k 2 \sinh{ \frac{\sigma s_i}{2\la}} .
\ee
This function is of course the generating function for the $n=0$ limit of $\frac{1}{n} \langle \rm{tr} B^{p_1}\cdots \rm{tr} B^{p_k}\rangle$ by expanding in the $s_i$'s. Selecting the coefficients of equal powers for
every $s_i$, for instance of $(s_1\cdots s_k)^3$,  we find 
\be \label {cube}
\lim_{n \to 0} \frac {1}{n} <(\rm{tr} B^3)^{4g-2}>  = \frac{3^{3g-2}2^{-2g} (6g-4)!}{\la^{6g-3}} \left(\begin{array}{cc}
4g-2 \\g
\end{array}\right)
\ee
all other powers of $ \lim_{n \to 0} \frac {1}{n} <(\rm{tr} B^3)^{k}>$ vanishing unless $k=2 \ (\rm{mod}4)$. 
This leads to the intersection number of the moduli of curves with one marked point.
Indeed , following Kontsevich, these numbers are given  by
\be \frac{1}{n}\log Z = \sum_l t_l\langle \tau_l\rangle
\ee
with 
\be
t_l = (-2)^{-(4l+2)/6}\prod_0^{l-1} (2m+1) (\frac{2}{\la})^{2m+1} \ee
from which the above result  (\ref{cube}) provides
\be \label{tau}
  <\tau_{3g - 2} >_{g} = \frac{1}{(24)^g g!} \hskip 4mm (g=0,1,2,...).
\ee
These numbers agree with the values of the intersection numbers
computed earlier by Kontsevich, Witten and others \cite{Kontsevich,Witten,Itzykson-Zuber2, LiuXu}.

Clearly the method allows one to compute more than that. For instance one can derive as well 

\be \label{B4}
\lim_{n \to 0} \frac {1}{n} <(\rm{tr} B^4)^{2p-1}>  = \frac { 2^{2p-2}}{\la^{4p-2}}(4p-3)! \left(\begin{array}{cc}
2p-1\\ p
\end{array}\right)
\ee
These vertices appear in the higher Airy fuctions and they are related to the
intersection numbers of Witten's top Chern class. \cite{Witten,Kontsevich}

\vskip 2mm
\section{Application to intersection numbers of top Chern class (p-spin curves)}
\vskip 2mm
  The partition function  for the p-th generalization of  Kontsevich's  Airy matrix-model 
is defined as 
\be\label{p}
 Z =\frac{1}{Z_0} \int dB {\rm exp}[ \frac{1}{p+1}\tr (B^{p+1} -\Lambda^{p+1}) - 
\tr (B - \Lambda)\Lambda^p]
\ee
normalized by the Gaussian part of the integrand (after the shift $B\to B+\Lambda$ which cancels the linear terms in $B$ of the exponent) : 
\be
Z_0 = \int dB {\rm exp}[ \sum_{j=0}^{p-1}\tr \frac{1}{2}\Lambda^j B \Lambda^{p-j-1}B]
\ee
The 'free energy', i.e. the logarithm of the partition function, 
is the generating function of the generalized intersection numbers $\langle \prod \tau_{m,j}\rangle$ for
 moduli of curves with 'spin' j \cite{Kontsevich,Jarvis}

\be\label{expansion}
F= \sum_{d_{m,j}} < \prod_{m,j} \tau_{m,j}^{d_{m,j}}> \prod_{m,j} \frac{t_{m,j}^{d_{m,j}}}{d_{m,j}!}
\ee
where \cite{Kontsevich}

\be\label{t's}
  t_{m,j}=(-p)^{\frac{j-p-m(p+2)}{2(p+1)} }\prod_{l=0}^{m-1}(lp+j+1)\tr \frac{1}{\Lambda^{mp+j+1}}
\ee
 According to Witten \cite{Witten2} the intersection numbers is given by 
\be
<\tau_{m_n,j_n} \cdots \tau_{m_n,j_n}>_g = \int_{\bar M_{g,n}} C_w(j_1,\cdots,j_n) \psi_1^{m_1}\cdots
\psi_n^{m_n}
\ee
with the condition with relates, for given $p$, the indices to the genus $g$ of the surface
\be
 (p+1)(2g - 2+n) = \sum_{i=1}^n (p m_i + j_i + 1).
\ee
The cohomology class $C_w(j_1,\cdots,j_n)$ is Witten's class (top Chern class), and 
$\psi_1$ is the first Chern class.
Witten  conjectured  that $F$ is the string solution of the p-th Gelfand-Dikii hierarchy
\cite{Witten2,Adler}. 
The partition function $Z$ of (\ref{p}) is a model for the (p,1)-quantum gravity,
which is equivalent to the (p-1)-matrix model.

 By retaining only  the leading term of the limit $n\rightarrow 0$ limit, we restrict ourselves to surfaces with one marked point. 
Then the expansion of (\ref{expansion}) is simply given by
\be
 {\rm lim}_{n\rightarrow 0} \frac{1}{n}F = \sum_{m,j} <\tau_{m,j}>_g t_{m,j}.
\ee
In the leading order of the limit $n\rightarrow 0$, the matrix $\Lambda$ is replaced by a scalar $\Lambda=\lambda\cdot I$.

  After a simple rescaling of  the Gaussian weight, we obtain, for instance in the case of p=3,
\be\label{p3}
  Z = \frac{1}{Z_0} \int dB {\rm exp}[\frac{1}{36} \tr B^4 + \frac{i}{3\sqrt{3}}\tr \Lambda B^3
-\frac{1}{2}\tr \Lambda^2 B^2]
\ee
\be
Z_0= \int dB {\rm exp }[-\frac{1}{2} \tr \Lambda^2 B^2]
\ee
In this  p=3 case, we obtain from (\ref{sh}),
\be
<\tau_{\frac{8g-5-j}{3},j}>_g=\frac{1}{(12)^g g!} \frac{\Gamma(\frac{g+1}{3})}{\Gamma(\frac{2-j}{3})}
\ee
where $j=0$ for $g=1,4,7,10,\cdots$ and $j=1$ for $g=3,6,9,\cdots$. For $g=2,5,8,\cdots$, the intersection numbers are
zero.

When the genus g is equal to three, the above formula gives
  $<\tau_{6,1}>_{g=3} = \frac{1}{(12)^3 3! 3}$ 
which agrees with the value obtained earlier by
Shadrin  \cite{Shadrin}.
In the calculation of the intersection numbers for p=3,  in addition to $< (\tr B^3)^p> $and $<(\tr B^4)^q> $   given hereabove in (\ref{cube}) and (\ref{B4}) , one needs to compute mixed averages of the type
$<(\tr B^4)^k (\tr B^3)^l> $. Such averages are indeed required if one deals with the expansion of (\ref{p3}), but they are also contained in the explicit formula (\ref{sh}) for the generating function. 

For general p, from (\ref{sh}),  we have for g=1, $<\tau_{1,0}>_{g=1} = \frac{p-1}{24}$.
To derive this result, we simply need to compute  $<\tr B^4>_{g=1}=1$ and $<(\tr B^3)^2>_{g=1}=3$ with
$t_{1,0}= - n/(p \lambda^{p+1})$.
In the case p=4, we have $<\tau_{6,0}>_{g=3} = \frac{9}{8^3 5!}$.

Thus we find that the expression of (\ref{sh}) for the $n \to 0$ limit of $U(s_1,\cdots,s_k)$  is a generating function for the 
intersection numbers for the moduli space of one-marked point, p spin, curves.
This provides thus a simple  algorithm for obtaining these numbers. 

We have restricted this article to surfaces with one marked point. 
To go beyond one marked point, the analysis
of higher orders in $n$ is required. For two marked points, we need the order $n^2$. 
However for genus one, we have two different kinds of intersection numbers, $<\tau_0 \tau_2>$ and
$<\tau_1^2>$  (here p=2). These two terms are distinguished by coupling to different combinations of the $t_l$ parameters, namely 
$t_0 t_2$ and $t_1^2$ ; those combinations have different $\Lambda$ dependence (\ref{t's}). Thus we need 
a matrix  $\Lambda$ which is no longer a multiple of the identity. If we go to higher orders in $n$ with simply for $\Lambda$ a multiple of the identity matrix,  we obtain the values of the sum
$<\tau_0\tau_2> +<\tau_1^2> = \frac{1}{12}$, instead of that of the  individual terms.
 We hope to be able to discuss higher marked point by extending the present approach to such cases.

\section{Conclusion}
The present article, using both duality and replica,  provides exact formulae for the intersection numbers of the moduli of curves on Riemann surfaces with one marked point and p-spin. The method may be extended, at least for low genera, to a higher number of marked points. It provides a relationship between Kontsevich's Airy matrix model and Okounkov's work of the edge of Wigner's semi-circle.

\newpage
\appendix {\bf{ Appendix A :  Gaussian averages in the presence of an external matrix source}}
\vskip 2mm
For the sake of completeness we reproduce here the main steps of the derivation given in \cite{BHc}. 
The probability distribution for Hermitian $n\times n$   matrices  is  thus
\be  P_A(B) = \frac{1}{Z} e^{-\frac{\la}{2} \rm{tr} B^2 + \rm{tr} AB} \ee
Let us first compute the one point function
\be U_A(s) = \frac{1}{n} \langle \rm {tr} e^{sB}\rangle .\ee
Using the Itzykson-Zuber formula \cite {Harish-Chandra, Itzykson-Zuber} to integrate out the unitary degrees of freedom  one obtains :
\be U_A(s) = \frac{1}{n} \frac{1}{Z\Delta(A)} \int \prod_1^n db_i \Delta(B)  e^{\sum_1^n(-\frac{\la}{2} b_i^2 + a_ib_i)} \sum_1^n e^{sb_i}\ee
in which $Z$ is a normalization constant fixed such as $U_A(0)=1$ and $\Delta(B)$ is the VanderMonde determinant of the eigenvalues $\Delta(B) = \prod_{i<j} (b_i-b_j)$. 
We now use repeatedly the trivial identity
\be \label{trivial} \int \prod_1^n db_i \Delta(B)  e^{\sum_1^n(-\frac{\la}{2} b_i^2 + a_ib_i)}  = C_ne^{\sum_1^n\frac{1}{2\la} a_i^2} \Delta(A) \ee
in which $C_n$ is a simple constant,  with $a_i$ replaced by $\tilde{a_i}= a_i +s\delta_{i,\alpha} $ in which $\alpha$  takes the successive values 1 to n.  This gives
\ba U_A(s)&& = \frac{1}{n}e^{\frac{s^2}{2\la}}\sum_1^n e^{sa_{\alpha}/\la} \frac{\Delta(\tilde{A_{\alpha}})}{\Delta(A)}\nonumber \\&& =\frac{1}{n}e^{\frac{s^2}{2\la}} \sum _{\alpha} e^{s a_{\alpha} /\la} \prod_{\beta \neq \alpha} \frac{ a_{\alpha}- a_{\beta} +s} { a_{\alpha}- a_{\beta}}  \ea
which may be replaced by the contour integral around the n points $(a_1, \cdots, a_n)$ 
\be U_A(s) =  \frac{1}{ns}e^{\frac{s^2}{2\la}} \oint \frac{du}{2i\pi} e^{su/\la} \prod_1^n \frac{ u+s-a_{\alpha}}{u-a_{\alpha}}.\ee

Note that this is an exact formula ; simplifies of course for the pure Gaussian case $a_{\alpha}=0$, although the derivation does require a source matrix in the intemediate steps. 

The derivation for the k-point functions follows exactly the same lines : the identity (\ref{trivial}) requires now an $\tilde a$ which carries k-indices.  The result of the calculation is similar : 
\ba && U_A(s_1,\cdots, s_k)= \frac{1}{n} \langle \rm {tr} e^{s_1B}\cdots \rm {tr} e^{s_k B} \rangle \nonumber \\&&=   \frac{1}{n s_1\cdots s_k }e^{\sum_1^k\frac{s_i^2}{2\la}}\oint \prod _{i=1}^k\frac{du_i}{2i\pi} e^{\sum_{i=1}^k s_iu_i /\la}\prod_{i<j} \frac{(u_i-u_j+s_i-s_j)(u_i-u_j)}{(u_i-u_j+s_i)(u_i-u_j-s_j)} \nonumber \\&& \times \prod_{i=1}^k\prod_{\alpha=1}^n (1+\frac {s_i}{u_i-a_{\alpha}}) .\ea
One can simplify the expression by noticing the Cauchy determinant identity
\be \det {\frac{1}{ x_i-y_j} }= (-1)^{n(n-1)/2} \ \frac {\prod_{i<j} (x_i-x_j)(y_i-y_j)}{\prod_{i,j} (x_i-y_j)} 
\ee
with $x_i = u_i+s_i, y_i= u_i$, namely
\be 
\det {\frac{1}{u_i+s_i-u_j} }=  \frac{1}{ s_1\cdots s_k }\prod_{i<j} \frac{(u_i-u_j+s_i-s_j)(u_i-u_j)}{(u_i-u_j+s_i)(u_i-u_j-s_j)} .
\ee
In the $n=0$ limit we need to consider only the  the connected part of $U_A$ (the disconnected ones vanishing with n, as is obvious from Feynman diagrams or explicit formulae)) which reads, which correspond to connected permutations in the expansion of the determinant.  Taking a vanishing source ($a_{\alpha} = 0$) the $n=0$ limit gives immediately (\ref{k-point}). 

\newpage
\appendix {\bf{Appendix B :  Duality}}
\vskip 2mm

Let us consider the Gaussian average, with a matrix source $A$, of  the product of characteristic determinants of the $N\times N$
Hermitian  random matrices
 \ba
F_k(\lambda_1,...,\lambda_k) &=& \frac{1}{Z_N}< \prod_{\alpha=1}^k {\rm det}(\lambda_\alpha - M) >_{A,M}\nonumber \\
&=& \frac{1}{Z_N}\int dM \prod_{\alpha=1}^k {\rm det}(\lambda_i\cdot {\rm I} - {\rm M}) e^{-\frac{1}{2}
\Tr (M-A)^2} 
\ea
in which $Z_N = \int dM exp(-1/2 \Tr M^2) =2^{N/2} (\pi)^{N^2/2}$ .
The unitary invariance of the measure allows us to assume, without loss of generality,  that $A$ is a diagonal matrix with eigenvalues $(a_1, \cdots,a_N)$. We introduce now $N\times k$ complex Grassmanian variables ( $i=1,2,\cdots, N$ and $ \alpha = 1,2, \cdots, k$)
 $(\bar{ \psi}_i^{\alpha},  \psi_i^{\alpha})$  ( $i=1,2,\cdots, N$ and $ \alpha = 1,2, \cdots, k$) with the normalization
\ba \int d\bar{\psi}d \psi \  \left(\begin{array}{cc}
1\\ \bar{\psi} \psi 
\end{array}\right) =  \left(\begin{array}{cc}
0\\ 1
\end{array} \right)
\ea
Then one may write

\be \prod_{i=1}^k {\rm det}(\lambda_i\cdot {\rm I} - {\rm M})= \int \prod d\bar{ \psi}_i^{\alpha} d \psi_i^{\alpha} e^{{\sum_{i,j=1}^N\sum_{\alpha = 1}^k } \bar{ \psi}_i^{\alpha} (\la_{\alpha} -M)_{ij},  \psi_j^{\alpha}}\ee

and perform the Gaussian average 
\be \frac{1}{Z_N} \int dM  e^{-\frac{1}{2}
\Tr M^2 +   \Tr M X} =  e^{\frac{1}{2}
\Tr X^2 } . \ee
We deal here with a matrix $X$ given by
\be X_{pq} = a_p\delta_{pq} - \sum_{\alpha = 1}^k  \bar{ \psi}_q^{\alpha}  \psi_p^{\alpha} \ee
and thus
\be \label{trace}\rm{Tr} X^2 = \rm{Tr} A^2 - 2 \sum_{p=1}^N \sum_{\alpha=1}^k a_p \bar{ \psi}_p^{\alpha}  \psi_p^{\alpha} -\sum_{\alpha, \beta = 1,\cdots,k} \gamma_{\alpha, \beta} \gamma_{\ \beta, \alpha} \ee
where
\be \gamma_{\alpha, \beta} = \sum_{i=1}^N  \bar{ \psi}_i^{\alpha}  \psi_i^{\beta} \ee
To make the notations more transparent let us denote by 'Tr' the trace on the initial space of $N\times N$ matrices and by 'tr' the trace in the new space of $k\times k$ matrices : the last term in  (\ref{trace}) for $\rm{Tr} X^2$  is thus $-\rm{tr} \gamma^2$, the minus sign being due to the anticommuting nature of the psi's. We then introduce an auxiliary Hermitian $k\times k$ matrix $\beta$ so that
\be e^{-\frac{1}{2}\tr \gamma^2 } = \frac{1}{Z_k} \int d\beta e^{-\frac{1}{2}
\tr \beta^2  + i \rm{tr} \gamma \beta}. \ee

This leads us to the representation
 \ba
F_k(\lambda_1,...,\lambda_k) &&=\frac{1}{Z_k} \int  \prod d\bar{ \psi}_i^{\alpha} d \psi_i^{\alpha}  \nonumber \\&& \times \int d\beta  e^{-\frac{1}{2}
\tr \beta^2  + i \rm{tr} \gamma (\beta-i\Lambda)}e^{-\sum_{i=1}^N \sum_{\alpha=1}^k a_i \bar{ \psi}_i^{\alpha}  \psi_i^{\alpha} }
 \ea
 in which $\Lambda$ is the $k\times k$ diagonal matrix with eigenvalues $(\lambda_1, \cdots, \lambda_k)$. 
 One can now easily perform the integration over the Grassmanian variables since the exponent is quadratic in those variables. This yields  the announced dual representation
\be \label{duality}  F_k(\lambda_1,...,\lambda_k) =\frac{1}{Z_k} \int d\beta  e^{-\frac{1}{2}
\tr \beta^2} \prod_{j=1}^N \det [( \la_{\rho}-a_j) \delta_{\rho, \sigma} + i\beta_{\rho, \sigma} ] \ee
in which the matrices are now $k\times k$. 

Remark : this duality takes  a more symmetric form if we chose  a pure imaginary diagonal matrix A, and if we shift the matrix $\beta$  by $i\Lambda$ in the final formula (\ref{duality}).  Then the duality  reads

\ba &&\frac{1}{Z_N}\int dM \prod_{\alpha=1}^k {\rm det}(\lambda_i\cdot {\rm I} - {\rm M}) e^{-\frac{1}{2}
\Tr (M-iA)^2} = \nonumber \\
 &&(-i)^{Nk} \frac{1}{Z_k} \int d\beta  \ e^{-\frac{1}{2}
\tr (\beta+i\Lambda)^2} \prod_{j=1}^N \det (a_j \delta_{\rho, \sigma} - \beta_{\rho, \sigma} ) 
 \ea

This duality may be extended to a two-matrix model when the measure is an exponential of a quadratic form in the two matrices $M$ and $M'$, namely $\exp{(a TrM^2 + b Tr M'^2 + c TrMM')}$.  

\newpage {\bf{ Appendix C: Contour integrals in the $n=0$ limit}}
 \vskip 2mm
We return to the contour integral (\ref{k-point}) which gives the $n=0$ limit of the k-point function. 
Let us consider for simplicity the case of the two-point function. We are then dealing with the sum of two integrals :
\ba\label{2-point} \lim_{n\to 0} U(s_1,s_2)  && = -
 e^{\frac{s_1^2+s_2^2}{2\la}} \oint \prod_1^2 \frac{du_i}{2i\pi} e^{\sum_1^2 (u_is_i/\la)}  [\log{(1+\frac{s_1}{u_1})}+\log{(1+\frac{s_2}{u_2})} ] \nonumber\\ &&\times \det \frac{1}{u_i+s_i-u_j}.
\ea
The continuation to $n=0$ requires to take contours in the $u_1$ and $u_2$ planes which circle around the respective cuts $[-s_1,0]$ and $[-s_2,0]$. We have to choose some well-defined contours on the two variables before we can write the integral (\ref{2-point}) as a sum of two integrals. For instance we choose and integral over a large contour in the $u_2$ plane , and close to the cut in the $u_1$ plane. The disconnected term $1/s_1s_2$ of the determinant gives a vanishing contribution. Thus
\ba U(s_1,s_2)=&&
 e^{\frac{s_1^2+s_2^2}{2\la}} \oint \prod_1^2 \frac{du_i}{2i\pi} e^{\sum_1^2 (u_is_i/\la)}  [\log{(1+\frac{s_1}{u_1})}+\log{(1+\frac{s_2}{u_2})} ]\nonumber \\ && \times\frac{1}{(u_1+s_1-u_2)(u_2+s_2-u_1)}
\ea
The second integral, the one which involves $\log{(1+\frac{s_2}{u_2})}$, vanishes for the choice of contours that we have made, since we can integrate over $u_1$ first and there are no poles inside the countour. For the first part, that with $\log{(1+\frac{s_1}{u_1})}$,  we integrate over $u_2$ first, pick up the two poles and find
\be U(s_1,s_2) = 
 e^{\frac{s_1^2+s_2^2}{2\la} }\frac{1}{\sigma}(\e^{-s_2^2/\la} - e^{ s_1s_2/\la}) \oint  \frac{du_1}{2i\pi} e^{u_1\sigma/\la} \log{(1+\frac{s_1}{u_1})} \ee
 with $\sigma = s_1+s_2$. Since 
 \be \oint  \frac{du_1}{2i\pi} e^{u_1\sigma/\la} \log{(1+\frac{s_1}{u_1})} = -\int_{s_1}^0 dx e^{x\sigma/\la}  = -\frac{\la}{\sigma} (1-e^{-s_1\sigma/\la}) \ee
 we end up with 
 \be
U(s_1,s_2) = 4\frac{\la}{\sigma^2} \sinh {\frac{\sigma s_1}{2\la}}\sinh {\frac{\sigma s_2}{2\la}}
\ee

This calculation may be repeated for the k-point function, although the combinatorics becomes heavy. 
This yields the final result
\be \label{final} \lim_{n\to 0}U(s_1,s_2,\cdots, s_k) = \frac{\la}{\sigma^2} \prod_1^k 2 \sinh{ \frac{ s_i \sigma}{2\la}} \ee
where $ \sigma = s_1+s_2+\cdots+ s_k$. 

It is straightforward to generalize this formula to the case of a non-zero external matrix source $A$ with eigenvalues $(a_1,\cdots, a_n)$. We assume that the density of eigenvalues
\be \rho (x) =\frac{1}{n} \sum_1^n \delta (x-a_i) \ee
has a finite limit $\rho_0 (x)$when $n$ goes to zero. Up to now we have dealt with $\rho_0(x) = \delta(x)$, but we could take other examples such as $n/2$ eigenvalues  equal to $+a$ and $n/2$ equal to $-a$. Then we would deal with $\rho_0(x) = \frac{1}{2} [\delta(x-a)+ \delta (x+a)]$. The derivation goes through now in the same way : we write 
\be 
\prod_1^n( 1+\frac{s}{u-a_\alpha}) = e^{n \int dx \rho(x) \log{ (1+ \frac{s}{u-x})}} 
\ee
which may be expanded when $n$ goes to zero provided, as we have assumed, that $\rho(x)$ has a limit. The result is then simply
\be \lim_{n\to 0}U_A(s_1,s_2,\cdots, s_k) = \lim_{n\to 0}U(s_1,s_2,\cdots, s_k) \int dx \rho_0(x) e^{x\sigma/\la} \ee
in which the first factor $U$ is given again by (\ref{final}).
\vskip 2mm

{\bf{Acknowledgments}} : We are grateful to Profs. Zagier and Kontsevich for their interest and suggestions.

\end{document}